# Disruption Precursor Onset Time Study Based on Semi-supervised Anomaly Detection

X. K. Ai[1], W. Zheng[1]*, M. Zhang[1]*, D.L. Chen[2], C. S. Shen[1], B.H. Guo[2], B.J. Xiao[2], Y. Zhong[1], N.C. Wang[1], Z.J. Yang[1], Z. P. Chen[1], Z.Y. Chen[1], Y. H. Ding[1], Y. Pan[1], and J-TEXT team[1]

[1] International Joint Research Laboratory of Magnetic Confinement Fusion and Plasma Physics, State Key Laboratory of Advanced Electromagnetic Engineering and Technology, School of Electrical and Electronic Engineering, Huazhong University of Science and Technology, Wuhan 430074, China

[2] Institute of Plasma Physics, Chinese Academy of Sciences, Hefei 230031, China

Corresponding author: W. Zheng, M. Zhang

E-mail: zhengwei@hust.edu.cn, zhangming@hust.edu.cn

**Fax:** +86-27-87793060

## Abstract

The full understanding of plasma disruption in tokamaks is currently lacking, and data-driven methods are extensively used for disruption prediction. However, most existing data-driven disruption predictors employ supervised learning techniques, which require labeled training data. The manual labeling of disruption precursors is a tedious and challenging task, as some precursors are difficult to accurately identify, limiting the potential of machine learning models. To address this issue, commonly used labeling methods assume that the precursor onset occurs at a fixed time before the disruption, which may not be consistent for different types of disruptions or even the same type of disruption, due to the different speeds at which plasma instabilities escalate. This leads to mislabeled samples and suboptimal performance of the supervised learning predictor. In this paper, we present a disruption prediction method based on anomaly detection that overcomes the drawbacks of unbalanced positive and negative data samples and inaccurately labeled disruption precursor samples. We demonstrate the effectiveness and reliability of anomaly detection predictors based on different algorithms on J-TEXT and EAST to evaluate the reliability of the precursor onset time inferred by the anomaly detection predictor. The precursor onset times inferred by these predictors reveal that the labeling methods have room for improvement as the onset times of different shots are not necessarily the same. Finally, we optimize precursor labeling using the onset times inferred by the anomaly detection predictor and test the optimized labels on supervised learning disruption predictors. The results on J-TEXT and EAST show that the models trained on the optimized labels outperform those trained on fixed onset time labels.

# 1 Introduction

Plasma disruption is a phenomenon that occurs in all tokamaks, which is a sudden loss of the plasma thermal energy and plasma current which will deposit a strong electromagnetic force and a large heat load to the tokamak device[1]-[3]. Precursors of plasma instability and disturbance are commonly observed before the occurrence of disruptions. Based on these precursors, a set of methods have been developed to predict disruption. When disruption precursors are detected, the disruption mitigation system (DMS) can be triggered to reduce the damage to the tokamak machine. Most current disruption prediction follows two approaches: physics-driven or data-driven. As the theory of plasma disruption in tokamaks is not fully understood, the development of a reliable physics-driven predictor is challenging[4]. Physics-driven predictors that used the locked mode signal for disruption prediction have been tested on JET[5]. Although the predictor has a certain extrapolation ability, its performance is inadequate for future high-power tokamaks (such as ITER).

Data-driven methods, such as support vector machines[6],[7], random forests[8],[9], and neural networks[10]-[13], have been utilized to develop disruption predictors based on supervised learning models. However, these methods require labeled data and a balanced dataset of positive and negative samples to ensure their performance. Many supervised learning disruption prediction predictors have been developed in EAST[14],[15], HL-2A[16],[17], JET[18]-[20], DIII-d[21],[22], ASDEX-U[23]-[25], NSTX[26] and JT-60U[27],[28]. Although these types of predictors show good performance, some predictors can achieve a success rate of over 90% and a false alarm rate of less than 10%. But they demand a massive amount of data for training, especially in obtaining positive samples. In the context of disruption prediction, obtaining positive samples requires a large number of disruption shots, which can cause significant damage to the machine. Therefore, future tokamaks, such as ITER, will only allow non-disruption shots, and only a very small number of disruption shots will occur. This leads to an extremely unbalanced dataset, mostly consisting of non-disruption shot data. Thus, the future disruption predictors for reactors will likely be trained on only or mostly non-disruption shots and will try their best to avoid disruption. However, for supervised learning methods, the unbalanced dataset may negatively impact the performance of the trained model[29],[30].

One additional challenge faced by supervised learning disruption predictors is the identification and labeling of disruption precursors. However, the underlying physical principles of plasma disruption remain unclear, making it difficult to determine the precursor onset time using physical rule-based algorithms. While experts may manually determine some precursor onset times, this process is time-consuming and cannot generate a large enough dataset. The survey of the previous citation on the supervised learning predictor[14]-[28] reveals that the mainstream labeling method is to assume all samples in a fixed period before disruption for all disruption shots as positive, while labeling samples in non-disruption shots as negative. For example, the assumed precursor sample period on JET is 40ms before plasma current quench time ($t_{CQ}$) in Cannas, B. et al. (2010)[31], 75ms on C-Mod and 400ms on DIII-D in Zhu, J. et al.

(2021)[32], and so on. The models of fixed period labeling method in these articles have achieved good performance. However, for different disruption types or even the same disruption type, there is no guarantee the time from the onset of disruption precursor to disruption is always the same. As different plasma instabilities escalate at different speeds, the assumed precursor onset time may differ from the actual onset time, leading to mislabeled time samples and contaminated training data that can severely hurt the performance of the disruption prediction predictor. To address this issue, a clustering method[33] has been studied for labeling disruption precursor samples, and the anomaly detection predictors presented in this paper offer another promising approach.

Training anomaly detection models only requires negative labeled samples, which are very easy to obtain form the non-disruption shots[34]-[37]. During the flat-top phase of non-disruptive shots, there is no disruption precursor present. When the model infers, the disruption precursor samples are classified as anomalous points, and a disruption alarm is issued. Anomaly detection disruption predictors address the issues of unbalanced positive and negative data samples and incorrectly labeled disruption precursor samples, making them suitable for deployment on future devices. In this paper, anomaly detection predictors based on different algorithms are established and tested on J-TEXT and EAST. One of which is a deep learning model based on autoencoder (AE). The rest are all traditional machine learning models based on: one-class support vector machine (OCSVM), k-nearest neighbor (KNN), and angle-based outlier detection (ABOD). We analyzed the precursor onset times inferred by these predictors and found that precursor onset times for different disruption shots are not necessarily the same, highlighting the need for further improvements in supervised learning labeling methods. Finally, we discovered that training the precursor sample training set labeled by the precursor onset time inferred by an anomaly detection predictor can improve the performance of supervised learning predictor.

This paper is organized as follows: Section 2 provides an overview of the predictor structure and model principles. Section 3 details feature selection and dataset preparation for training the models. Section 4 analyzes the performances and precursor onset time. In section 5 precursor onset times inferred by anomaly detection predictor are utilized to optimize precursor labels, and the supervised model trained with optimized labels is compared with models trained with the mainstream labeling method.

## 2 Design of anomaly detection predictor

In this work, high-performance disruption predictors based on anomaly detection are the bases for the precursor onset time study. The structure of anomaly detection predictor is shown in Figure 1. All signals of a time sample in a discharge are formed into a feature vector. These time sequences of feature vectors constitute a shot. Anomaly detection predictor is only trained by normal samples from non-disruption shots. During prediction, samples of predicting shot are fed to the predictor in chronological order to evaluate and return predicted labels. Normal samples are labeled -1, and disruption precursor samples are labeled 1. An alarm is triggered when the label changes from -1 to 1, which indicates the presence of a disruption precursor. The moment at which the alarm is issued corresponds to the disruption precursor onset time ($t_{onset}$) that is inferred by the model. However, in practice, false alarms may be triggered due to noisy signals. To mitigate such false alarms, the alarm is issued only when the predicted label is 1 for N consecutive times during the disruption prediction, which is referred to as the decision length ($dl$). The decision length can be optimized as a hyperparameter during both model training and inference to improve the accuracy and reliability of the predictor. And disruption precursor onset time can be calculated:

$$t_{onset} = t_{alarm} + dl \cdot t_r \qquad (2\text{-}1)$$

The $t_{onset}$ is the time from precursor onset time to $t_{CQ}$. The $t_{alarm}$ is the time from alarm issued time to $t_{CQ}$. $t_r$ is time resolution, which is the time interval between two consecutive samples.

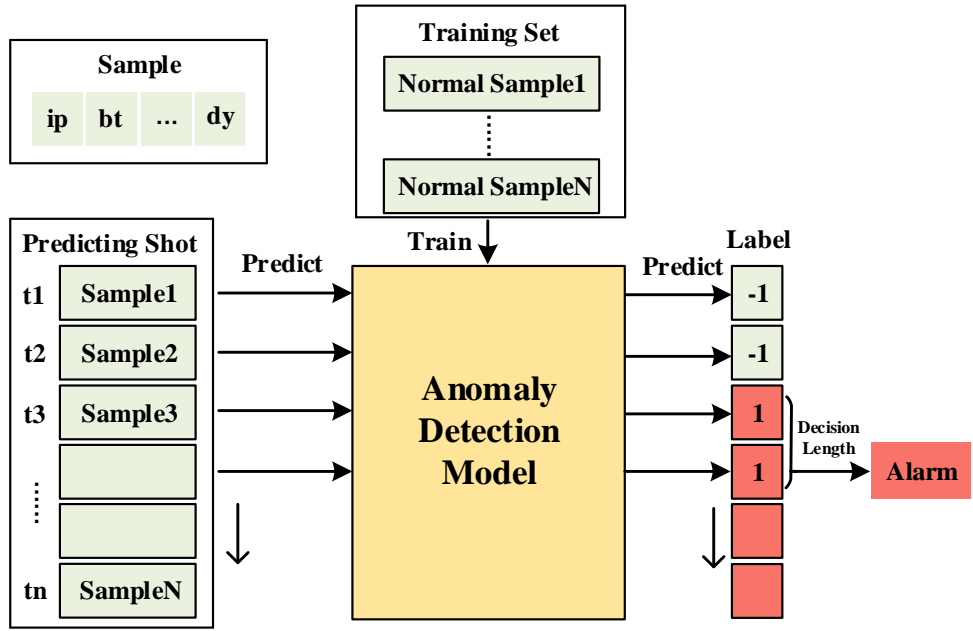

Figure 1 Schematic of anomaly detection predictor. In training mode, the training set composed of normal samples is imported into the predictor. In prediction mode, samples of predicting shot are imported into the predictor in chronological order to evaluate and return predicted labels. When the number of consecutive abnormal labels reaches the decision length, the predictor issues a disruption alarm.

In this paper, four anomaly detection models based on different algorithms are established, and precursor onset times inferred by these models based on different principles are analyzed. The algorithms of these models are based on AE, OCSVM, KNN, and ABOD.

## Autoencoder Anomaly Detection

Autoencoder is a deep learning algorithm based on neural network. The algorithm compresses and reconstructs the input samples, and its structure is divided into two parts: encoder and decoder. Given the input space $X \in \chi$ and the feature space $h \in \mathcal{F}$. The autoencoder solves the two mappings $f$ and g to minimize the reconstruction error of the input features[38],[39].

$$f : \chi \to \mathcal{F}, \ g : \mathcal{F} \to \chi \qquad (2\text{-}2)$$

$$MSE = ||X - g[f(X)]||^2 \qquad (2\text{-}3)$$

$$f, g = \arg\min_{f,\, g} MSE \qquad (2\text{-}4)$$

When utilizing autoencoders for anomaly detection, the model is trained only with normal samples. The model has good reconstruction for normal samples and poor reconstruction for disruption precursor samples. Mean squared error ($MSE$) measures the similarity between input samples and reconstructed samples. For normal sample, the $MSE$ is small; for disruption precursor samples, the $MSE$ is large. A $MSE$ threshold is delineated to determine whether the inferred sample is abnormal.

**One-Class Support Vector Machine**

OCSVM algorithm maps the data samples to the high-dimensional feature space through the kernel function so that it has better aggregation, and then in feature space solving an optimal hyperplane includes normal samples compactly. If the sample point is inside the hyperplane, it is decided as normal, and if it is outside the hyperplane, it is decided as abnormal. The distance from the sample point to the hyperplane is considered an anomaly score to measure the anomaly degree of sample. Let $\{x_i, i = 1, \ldots, N\}$ are the training instances in the input space X, where $N$ is the number of observations, and X is a set of normal data. the hyperplane can be formulated as:

$$f(x) = \mathrm{sgn}\left((\omega \cdot \phi(x_i)) - \rho\right) = \mathrm{sgn}\left(\sum_{i=1}^{n} \alpha_i K(x, x_i) - \rho\right) \qquad (2\text{-}5)$$

Where $\alpha$ is the Lagrange multiplier. $\omega$ and $\rho$ are the target optimization problem variables. And $\phi$ maps the original data into feature space, which is a higher dimensional space to differentiate the non-separable data sets linearly, and $K(x_i, x_j) = \phi(x_i) \cdot \phi(x_j)$ is the kernel function. A more detailed illustration of this method can be found in Schölkopf et al. (2001)[40].

**K-Nearest Neighbor Anomaly Detection**

KNN algorithm can be used for anomaly detection[41]. This algorithm is based on sample density measurement. Plasma is constrained by the device during normal discharge. The corresponding normal sample points are concentrated in the feature space and confined to a specific area[42], which are less distant from nearby points. When the disruption precursor appears, the plasma gradually loses its restraint as the disruption precursor develops. The corresponding precursor sample points gradually move away from the normal sample cluster in feature space[42]. Due to the wide variety of disruption types, the distribution of precursor sample points is sparse and the distance from nearby points is larger. a threshold of distance can be found to separate precursor samples from normal samples. Euclidean distance is used as the anomaly score. Let $x = (x^{(1)}, x^{(2)}, \cdots, x^{(n)})^{\mathrm{T}}, x \in \chi$ are the training instances $x$ in the input space $\chi$, where $n$ is the number of features. $x_i, x_j \in \chi$, The Euclidean distance $L_2$ between $x_i$ and $x_j$ is:

$$L_2(x_i, x_j) = \left(\sum_{l=1}^{n} \left|x_i^{(l)} - x_j^{(l)}\right|^2\right)^{\frac{1}{2}} \qquad (2\text{-}6)$$

### Angle-Based Outlier Detection

ABOD algorithm is also based on sample density measurement. This algorithm compares the angles between pairs of distance vectors to other points to distinguish whether sample points are outliers. For sample point in the cluster, there will be data points in each direction, and the angle of the formed vector is also in each direction. Therefore, the angle difference between the different vectors formed by the points in the cluster and other points is very large, and the variance of the angle sequence is large. For points far away from the cluster, the angle difference between the formed vectors will be far smaller than the points in the cluster, the angle of the vector has a certain direction, and the variance of the angle sequence is also far smaller than that of the points in the cluster. a threshold of variance can be found to separate abnormal points from normal points. Given a database $D \subseteq R^d$, point $A, B, C \in D$. The angle-based outlier factor $\text{ABOF}(A)$ is the variance over the angles between the difference vectors of $A$ to all pairs of points in $D$:

$$\text{ABOF}(A) = VAR_{B,C \in N_k(A)} \left( \frac{\langle \overrightarrow{AB}, \overrightarrow{AC} \rangle}{\|\overrightarrow{AB}\|^2 \cdot \|\overrightarrow{AC}\|^2} \right) \qquad (2\text{-}7)$$

When making an abnormal decision for each point, the model needs to count all the points in the data set, which is very time-consuming. So only the nearest k points are calculated, which greatly reduces the time-consuming calculation of the model[43].

## 3 Data preprocessing

Data preprocessing is an important part of disruption prediction experiments. In this section data preprocessing both on J-TEXT and EAST is described. Section 3.1 describes the selection of diagnostic signals and feature extraction. Section 3.2 describes the dataset preparation for anomaly detection experiments and supervised learning experiments.

## 3.1 Feature selection

In our previous study, we found that the physic-guided feature extractor can effectively improve the disruption prediction performance[44]. In this paper, four types of features extracted by physic-guided are also selected, which are MHD instabilities related, radiation related, density related, and basic plasma control system (PCS) signals. On J-TEXT, selected features are basically consistent with those selected by IDP-PGFE[44], and only temporal and special features extracted from the CIII radiation array are added. The EAST has a divertor configuration with large elongation, so the method of calculating its mode number by the Mirnov probe can't be consistent with J-TEXT, this feature type is not considered in this paper for the time being. The arrays for measuring radiation signals on EAST are SXR and PXUV respectively. The input signals selected on the J-TEXT and EAST are shown in Table 1. The features used only on J-text are highlighted in the second column.

Table 1 Descriptions and symbols of all the features

| Types of features | Relation to disruption | Symbol |
|---|---|---|
| MHD instabilities related | 2/1 magnetic island growth; Multi-magnetic island overlaps; Locked mode | Mir_abs<br>Mir_fre<br>Mir_VV<br>mode_number_m (J-TEXT only)<br>mode_number_n (J-TEXT only)<br>n=1 amplitude |
| Radiation related | Temperature hollowing; Edge cooling; | P_rad (J-TEXT only)<br>SXR_core<br>CIII<br>CIII_array_kurt(skew,var) (J-TEXT only)<br>AXUV (pxuv)_array_kurt (skew, var)<br>SXR_array_kurt (skew, var) |
| Density related | Density limit | $n_{e0}$<br>FIR_array_kurt (skew, var) (J-TEXT only)<br>sum_$n_e$ (J-TEXT only) |
| Basic PCS signals | Plasma out of control; | $B_t$, $I_p$, vl<br>$I_{hf}$, $I_{vf}$,<br>$I_{hf}\_diff$, $I_{vf}\_diff$, ip_diff<br>$d_r$, $d_z$ |

## 3.2 Dataset preparation

For the J-TEXT dataset, the discharges in the shot range 1045962~1066648 from 2017 to 2019 are selected. The discharges with complete diagnostic signals are randomly picked for experiments. Anomaly detection predictors are trained only with samples from non-disruption shots, so 300 non-disruption discharges are selected as the training set. The size of validation set and test set is shown in Table 2. Supervised learning predictors used for the comparison of labeling methods are also trained with precursor samples of disruption shots, 225 disruption discharges are selected as the positive training set. The other shot sets are the same as those of anomaly detection predictors. All discharges are split into sample points from the plasma current flattop to $t_{CQ}$ per 0.1ms.

For the EAST dataset, we get the discharges in the shot range 54000~97000. The discharges with complete diagnostic signals are randomly picked for experiments. The training set of anomaly detection predictors consists of 400 non-disruption discharges. The size of validation set and test set is also shown in Table 2. The durations of discharges on J-TEXT are within 1s. On EAST, the shot durations are mostly 7s-30s, and some even exceed 100ms. And all discharges are split into sample points from the plasma current flattop to $t_{CQ}$ per 1ms. For supervised learning predictor, to keep the ratio of positive and negative samples of the training set between the two devices similar, 1000 disruption discharges are selected as the positive training set of supervised learning predictor on EAST. The other shot sets are the same as those of anomaly detection predictors.

Table 2 Dataset split on J-TEXT and EAST device

| | Non-disruption [J-TEXT] | Disruption [J-TEXT] | Non-disruption [EAST] | Disruption [EAST] |
|---|---|---|---|---|
| No. training shots | 300 | 225 | 400 | 1000 |
| No. validation shots | 80 | 80 | 100 | 100 |
| No. test shots | 110 | 110 | 200 | 200 |

# 4 Performance and precursor onset time analysis

This section discusses the performance of disruption predictors based on four different anomaly detection algorithms: AE, OCSVM, KNN, and ABOD. In order to verify the efficacy of each predictor, both J-TEXT and EAST datasets were utilized. Section 4.1 focuses on the prediction performance of each predictor. Section 4.2 studies the precursor onset times inferred by those predictors.

## 4.1 Prediction Performance

First, we assess whether each predictor has the ability to distinguish between normal shots and disruption shots. As demonstrated in Section 2, anomaly detection predictors can provide an anomaly score for each sample. A sequence of anomaly scores can be obtained for each discharge. Figure 2 and Figure 3 illustrate the normalized anomaly score sequences for all shots in the J-TEXT and EAST test sets, respectively. The time 0 of the x-axis represents the plasma current quenching time ($t_{CQ}$). And discharge durations of EAST test set shots are much longer than 1s, so the x-axis is taken as 1s before $t_{CQ}$. In these figures, the green and red lines represent the anomaly score curves for non-disruption and disruption shots, respectively. The opacity of each trace is very low, so the darker part indicates that the traces are more concentrated. For non-disruption shots, the green line is very dark in the low-value area from beginning to end, it indicates that the inferred anomaly scores of non-disruption shots are very low. For disruption shots, the closer to $t_{CQ}$, the higher the red line and the darker the color. The farther away from $t_{CQ}$, the lower the red line and submerged in green. This shows that the anomaly scores during the normal period of the disruption shots are low and similar to the anomaly scores of non-disruption shots. These results show that no matter whether on J-TEXT or EAST, anomaly detection predictors based on AE, OCSVM, KNN, and ABOD algorithms have the ability to distinguish between non-disruption shots and disruption shots before $t_{CQ}$.

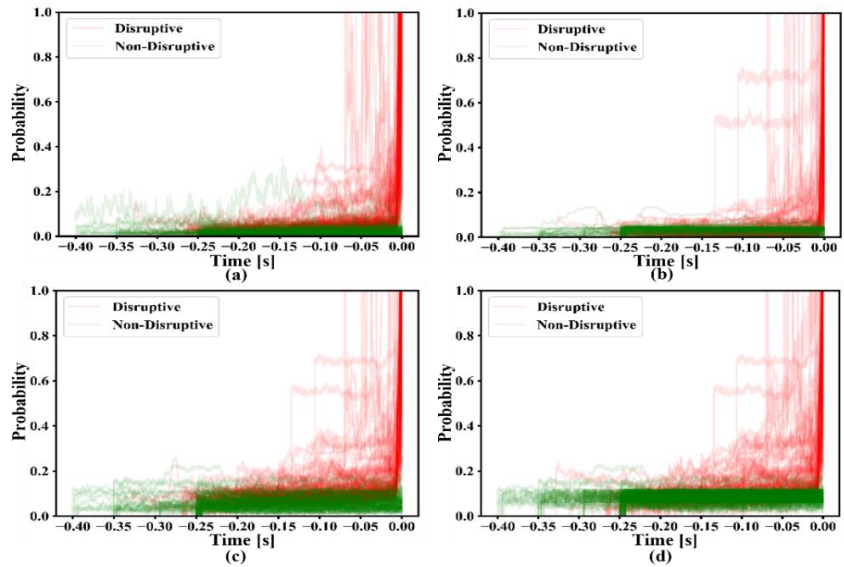

Figure 2 Anomaly score of test set shots on J-TEXT. (a) AE, (b) OCSVM, (c) KNN, (d) ABOD

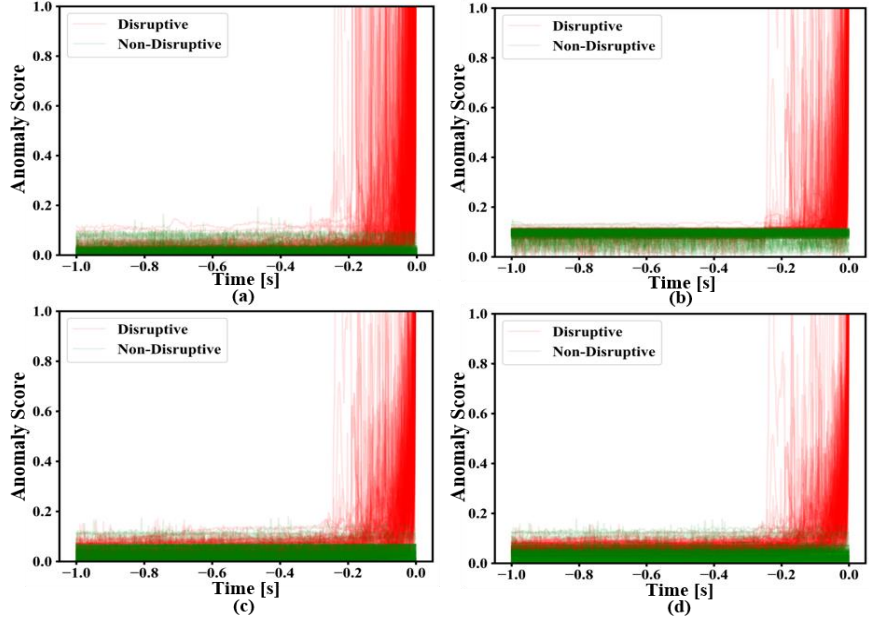

Figure 3 Anomaly score of test set shots on EAST. (a) AE, (b) OCSVM, (c) KNN, (d) ABOD

In disruption prediction, True positive (TP) refers to a successfully predicted disruption discharge. False positive (FP) refers to non-disruption discharges predicted as disruptions, also known as false alarms. True negative (TN) refers to non-disruption discharge predicted correctly. False negative (FN) refers to a disruption discharge not predicted, also known as the miss alarm and tardy alarm. The metrics used to evaluate the performance of disruption predictors often include accuracy, true positive rate (TPR), and false positive rate (FPR). They are calculated as follows:

$$Accuracy = \frac{TP + TN}{TP + FP + TN + FN} \tag{4-1}$$

$$TPR = \frac{TP}{TP + FN} \tag{4-2}$$

$$FPR = \frac{FP}{FP + TN} \tag{4-3}$$

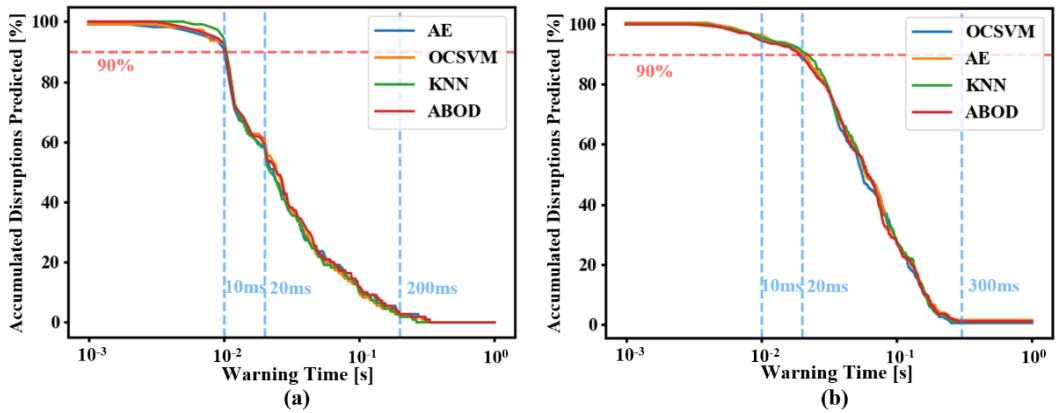

Figure 4 The accumulated percentage of disruption predicted versus warning time for AE, OCSVM、KNN, and

ABOD predictors. (a) On the J-TEXT device, for AE, KNN and ABOD predictors, the optimal thresholds obtained by hyperparameter search are 0.09, 0.13, 0.15, and decision lengths ($dl$) are 3, 6, 4 respectively. The OCSVM model is a binary classification model which does not need a threshold. For OCSVM predictor, $dl = 3$. (b) On the EAST device, for AE, KNN and ABOD predictors, the optimal thresholds obtained by hyperparameter search are 0.07, 0.10, 0.11, and decision lengths ($dl$) are 2, 3, 3 respectively. For OCSVM predictor, $dl = 3$.

For disruption predictor, it is necessary to reserve sufficient time for the MGI system to respond, which is referred to as the minimal warning time. In evaluating predictor performance, alarms issued within the minimal warning time before $t_{CQ}$ are considered tardy alarms and are not counted as true positives. Figure 4 illustrates the accumulated percentage of disruption predicted versus warning time for AE, OCSVM、KNN, and ABOD predictors on J-TEXT and EAST. It is worth noting that the output traces of anomaly detection predictors on J-TEXT and EAST are remarkably similar, indicating that the distributions of alarm times inferred by each anomaly detection predictor on the test set are highly consistent, which will be further expounded in Section 4.2. At the same prediction rate, such as 90% in Figure 4, the warning time on J-TEXT is shorter than that on EAST. Since J-TEXT is a small-sized tokamak with a relatively smaller time scale for disruption to take place[45],[46], the warning time is shorter than that for EAST[47],[48] or other large and medium sized tokamaks. Consequently, the selection of the minimal warning time is set to be 10ms on J-TEXT and 20ms on EAST. In a few cases, the alarm is issued in the relatively unstable region of the normal operation of disruption shot, which is very far from the $t_{CQ}$. Although the disruption shot is correctly predicted, it is deemed as a lucky guess. Thus, an upper limit of alarm time is established for evaluating model performance. If the alarm time exceeds the upper limit, the alarm is regarded as the premature alarm and will not be counted as true positive. In previous works, the upper limit of disruption precursor period on EAST is set as 300ms before disruption in Guo, B.H. et al. (2021)[15] and 500ms before disruption in Zhu, J. et al. (2021)[32]. In this paper, the upper limit of alarm time on EAST is set to be 300ms, while the upper limit of alarm time on J-TEXT is 200ms. Accordingly, the performance of all predictors on the J-TEXT and EAST test sets under these criteria is summarized in Table 3 and Table 4, respectively.

Table 3 Performance of AE, OCSVM, KNN, ABOD predictors on J-TEXT

| Predictor | Accuracy [%] | TPR [%] | FPR [%] |
|---|---|---|---|
| AE | 90.91 | 89.09 | 7.27 |
| OCSVM | 90.45 | 91.82 | 10.91 |
| KNN | 90.91 | 90.91 | 9.09 |
| ABOD | 89.10 | 90.00 | 11.82 |

Table 4 Table. 4 Performance of AE, OCSVM, KNN, ABOD predictors on EAST

| Predictor | Accuracy [%] | TPR [%] | FPR [%] |
|---|---|---|---|
| AE | 89.50 | 88.00 | 9.00 |
| OCSVM | 89.25 | 89.50 | 11.00 |
| KNN | 90.25 | 90.00 | 9.50 |
| ABOD | 89.75 | 87.50 | 8.00 |

In this study, we also investigate the feasibility of using the predictors for real-time disruption prediction. Table 5 displays the time consumption for inferring a sample by each predictor. The result demonstrates that KNN and ABOD predictors spend much more time in inference than AE and OCSVM predictors. This is because KNN and ABOD are models based on the sample density measurement. When inferring a sample, they compare sample point with a large number of points around it to evaluate the sparsity of its nearby position in the feature space. As a result, it involves substantial computation. In contrast, AE and OCSVM use training data to derive decision functions, which can be readily applied to input samples for inference. The inference times for AE and OCSVM predictors are significantly shorter, allowing for real-time disruption prediction. While KNN and ABOD predictors may encounter challenges in real-time application, they can still be utilized to verify the reliability of the precursor onset time inferred by the anomaly detection predictor in section 4.2.

Table 5 Time consumption for inferring a sample

| Model | AE | OCSVM | KNN | ABOD |
|---|---|---|---|---|
| Computing Time [s] | $4.55 \times 10^{-5}$ | $5.96 \times 10^{-5}$ | $4.30 \times 10^{-2}$ | $3.1 \times 10^{-2}$ |

## 4.2  Disruption precursor onset time study

The anomaly detection predictors with good performance can correctly predict most disruptions, and the false alarm rate is also very low, so it has the ability to identify disruption precursors. On this basis, their algorithms are completely different, and there is no manually labeled disruption precursor for training. Therefore, by comparing whether their inferred precursor onset times are similar, we can speculate whether they identify the disruption precursors according to the same rules. This section discusses whether precursor onset time inferred by anomaly detection predictor is reliable. This research includes two aspects: whether precursor onset times inferred by different anomaly predictors agree with each other and whether precursor onset time inferred by the anomaly detection predictor could represent the real precursor onset time. This section also considers tardy alarm shots, as in some cases the real disruption precursor could appear very close to the $t_{CQ}$. Figure 5 shows $t_{alarm}$ distributions of AE, OCSVM, KNN, ABOD predictors on the J-TEXT disruption shot test set. Figure 6 shows $t_{alarm}$ distributions of AE, OCSVM, KNN, ABOD predictors on the EAST disruption shot test set. Two figures show that $t_{alarm}$s of different disruption shots are not necessarily the same. $t_{onset}$ can be calculated by Formula 3-1, and the decision length for each model is fixed. So this reflects that precursor onset times of different disruption shots are not necessarily the same, which means that the fixed period labeling method is not optimal. Interestingly, $t_{alarm}$s are mainly distributed within 30ms in the J-TEXT disruption shot test set, while $t_{alarm}$s are mainly distributed above 30ms in the EAST disruption shot test set. In section 4.1, the maximum decision length time is 0.6ms among models on J-TEXT and 3ms among models on EAST. The result reflects that the precursor durations of disruption shots on the EAST device are generally longer than those on the J-TEXT device.

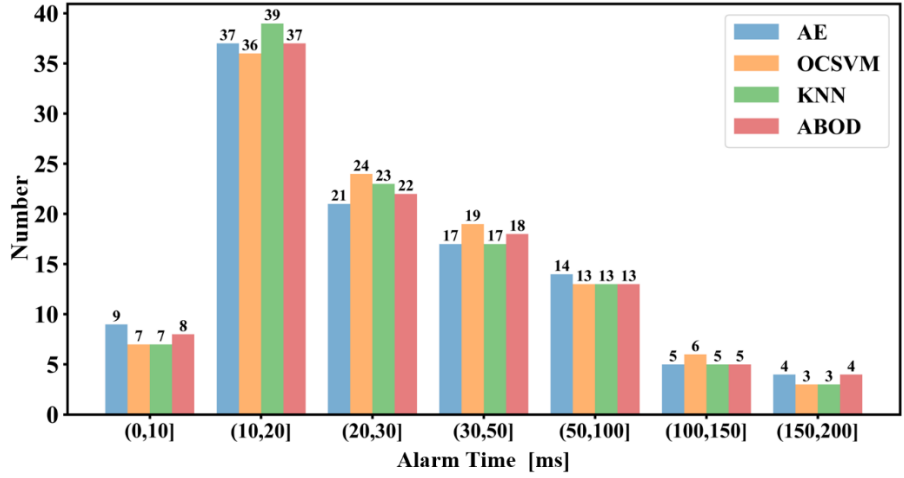

Figure 5 $t_{alarm}$ distributions of AE, OCSVM, KNN, ABOD predictors on J-TEXT disruption shot test set, the upper limit of $t_{alarm}$ is 200ms.

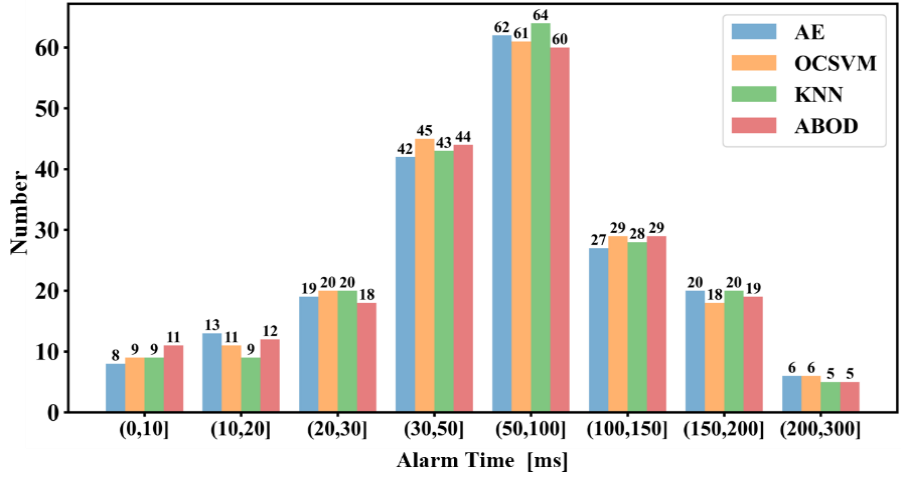

Figure 6 $t_{alarm}$ distributions of AE, OCSVM, KNN, ABOD predictors on EAST disruption shot test set, the upper limit of $t_{alarm}$ is 300ms.

Then we study whether $t_{onset}$s inferred by different predictors is similar for the same disruption shot. For the same disruption shot, AE, OCSVM, KNN, ABOD predictors infer four precursor onset times, and the $\Delta t$ is obtained by making a difference between four times. For example, for the same disruption shot, the difference between the AE predictor and the OCSVM(OC) predictor of inferred precursor onset time is $\Delta t_{AE-OC}$.

$$\Delta t_{AE-OC} = t_{onset,AE} - t_{onset,OC} \quad (4\text{-}4)$$

The $t_{onset}$s of AE and OCSVM predictors on J-TEXT are compared in Figure 7. Figure 7 (a) shows $t_{onset}$ distributions on J-TEXT disruption shot test set, while Figure 7 (b) shows $\Delta t_{AE-OC}$ distribution. And Figure 8 shows the comparison of $t_{onset}$s predicted by AE and OCSVM predictors on the EAST disruption shot test set. Both on the J-TEXT test set or the EAST test set, $t_{onset}$ distribution of AE or OCSVM is relatively discrete, but the majority of $\Delta t_{AE-OC}$s are close to 0. This reflects that the AE and OCSVM predictors infer similar precursor onset times for most disruption shots in the test set. It is also found that other $\Delta t$ distributions between the 4 predictors of AE, OCSVM, KNN, ABOD have the same as the $\Delta t_{AE-OC}$ distribution. All $\Delta t$ distributions show that Δ$t$s are mostly distributed around 0. Therefore, the disruption precursor onset times inferred by all anomaly detection predictors are roughly the same.

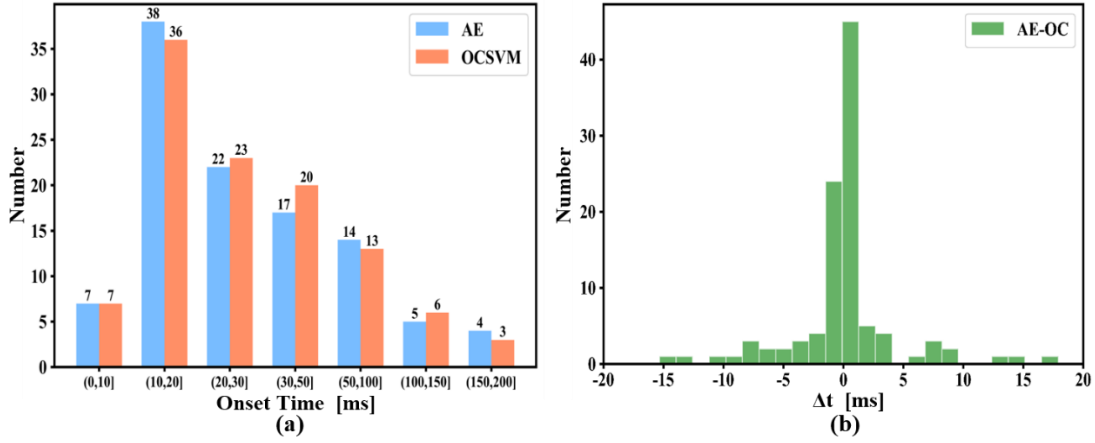

Figure 7 $t_{onset}$ comparison between AE, OCSVM(OC) predictors on J-TEXT disruption shot test set (a) $t_{onset}$ distributions of AE, OCSVM predictors (b) $\Delta t_{AE-OC}$ distribution

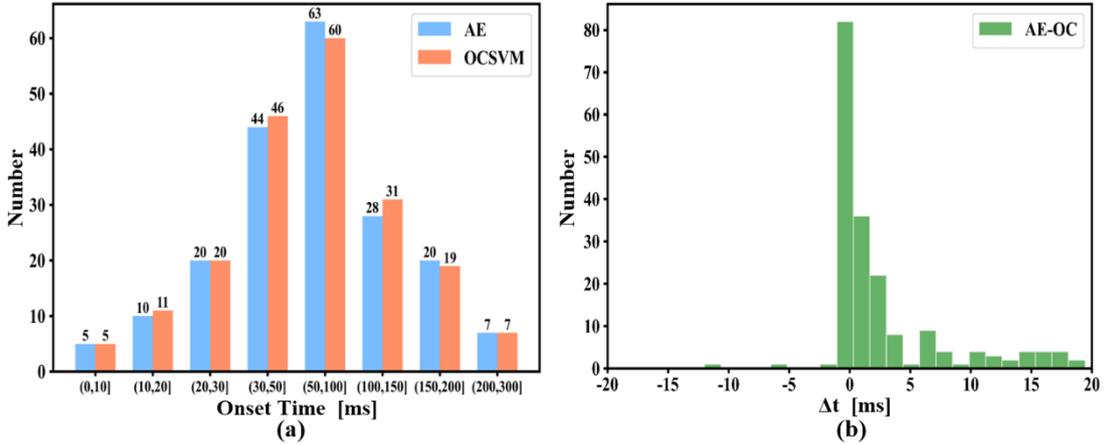

Figure 8 $t_{onset}$ comparison between AE, OCSVM(OC) predictors on EAST disruption shot test set (a) $t_{onset}$ distributions of AE, OCSVM predictors (b) $\Delta t_{AE-OC}$ distribution

Moreover, Table 6 calculates the mean value and standard deviation of $t_{onset}$s inferred by both two predictors for comparison in the J-TEXT disruption shot test set, as well as the mean and standard deviation of |Δ$t$|s. When calculating the mean and standard deviation, a small number of large outliers will significantly affect the calculated mean or standard deviation, thereby affecting the statistical results. The engineering error is caused by the mismatch of selected thresholds among some predictors. In order to eliminate this effect, when comparing $t_{onset}$s of two predictors in Table 6, the shots of |Δ$t$| > 50ms and premature alarm are removed. According to statistics, the number of shots with |Δ$t$| > 50ms and premature alarm of each predictor combination accounts for a very small proportion in the J-TEXT disruption shot test set. All proportions are between [6.36%, 9.09%], and the average is 8.36%. Table 7 is $t_{onset}$ comparison between those predictors in the EAST disruption shot test set. The shots of |Δ$t$| > 50ms and premature alarm are also removed. All proportions are between [2.54%, 5.08%], and the average is 4.39%. As seen from Figure 5 and Figure 6 that

$t_{onset}$ of different disruption shots are not necessarily the same on J-TEXT or EAST, so the standard deviations of $t_{onset}$s inferred by both comparative predictors are larger. However, the standard deviation of $|\Delta t|$ is significantly smaller than the standard deviations of $t_{onset}$s inferred by both comparative predictors. This phenomenon can reflect the similarity of $t_{onset}$s inferred by the two compared models. In summary, different anomaly detection predictors calculate similar precursor onset times for the same disruption shot.

Table 6 $t_{onset}$ comparison between predictors of AE, OCSVM(OC), KNN(KN), ABOD(AB) in J-TEXT disruption shot test set

| Item / Model | Mean[ms] | Std[ms] | Item / Model | Mean[ms] | Std[ms] | Item / Model | Mean[ms] | Std[ms] |
|---|---|---|---|---|---|---|---|---|
| AE | 39.51 | 42.70 | AE | 40.20 | 40.15 | AE | 39.24 | 40.51 |
| OCSVM | 40.62 | 41.54 | KNN | 39.15 | 41.16 | ABOD | 39.75 | 40.97 |
| $|\Delta t|_{AE,OC}$ | 4.547 | 8.187 | $|\Delta t|_{AE,KN}$ | 3.859 | 7.629 | $|\Delta t|_{AE,AB}$ | 4.318 | 7.768 |
| OCSVM | 43.55 | 39.85 | OCSVM | 43.03 | 46.01 | KNN | 41.65 | 42.87 |
| KNN | 41.12 | 40.23 | ABOD | 41.82 | 45.02 | ABOD | 42.07 | 41.43 |
| $|\Delta t|_{OC,KN}$ | 3.257 | 6.238 | $|\Delta t|_{OC,AB}$ | 5.368 | 8.833 | $|\Delta t|_{KN,AB}$ | 3.109 | 6.374 |

Table 7 $t_{onset}$ comparison between predictors of AE, OCSVM(OC), KNN(KN), ABOD(AB) in EAST disruption shot test set

| Item / Model | Mean[ms] | Std[ms] | Item / Model | Mean[ms] | Std[ms] | Item / Model | Mean[ms] | Std[ms] |
|---|---|---|---|---|---|---|---|---|
| AE | 74.71 | 55.17 | AE | 70.53 | 55.57 | AE | 75.94 | 54.67 |
| OCSVM | 71.33 | 52.05 | KNN | 70.24 | 54.96 | ABOD | 72.40 | 52.26 |
| $|\Delta t|_{AE,OC}$ | 6.155 | 9.999 | $|\Delta t|_{AE,KN}$ | 2.594 | 6.233 | $|\Delta t|_{AE,AB}$ | 4.005 | 8.524 |
| OCSVM | 75.33 | 52.24 | OCSVM | 74.66 | 51.76 | KNN | 77.04 | 54.21 |
| KNN | 70.17 | 54.71 | ABOD | 76.39 | 52.47 | ABOD | 74.75 | 52.43 |
| $|\Delta t|_{OC,KN}$ | 5.484 | 9.369 | $|\Delta t|_{OC,AB}$ | 2.818 | 6.380 | $|\Delta t|_{KN,AB}$ | 3.296 | 7.350 |

Next, we study whether the precursor onset time inferred by the anomaly detection predictor can represent the real one. Figure 9 shows diagnostic signals with disruption precursors of the disruption shot 1052267 on J-TEXT device and $t_{onset}$s inferred by AE, OCSVM, KNN, ABOD predictors which are marked with vertical lines of different colors. The time 0 of the x-axis represents $t_{CQ}$. $Mirnov_p$ is one channel from polar Mirnov array on J-TEXT, and Figure 9 (b)(c) show the frequency variation and amplitude variation of $Mirnov_p$. The frequency and amplitude of the Mirnov signal remained basically unchanged during the normal discharge period, and the frequency and amplitude decreased during the disruption precursor period, and finally mode-locked accrued. The turning point at which the frequency and amplitude start decreasing is a disruption precursor. And $t_{onset}$s inferred by all predictors are distributed closely around the turning point. Moreover, Figure 9 (d)-(e) show plasma electron density, skewness of soft x-ray profile, and kurtosis of CIII impurity profile on J-TEXT. For those signals, the waveform on the right side of the $t_{onset}$ markers change significantly compared to the waveform on the other side. Figure 10 shows the $t_{onset}$s inferred by anomaly detection predictors to disruption shots on EAST are also near the turning points of the corresponding diagnostic signals, which are also close to the turning points of the corresponding diagnostic signals. Therefore, we conclude that the precursor onset time inferred by the anomaly detection predictor is reliable and agrees with the disruption precursor recognized by human experts.

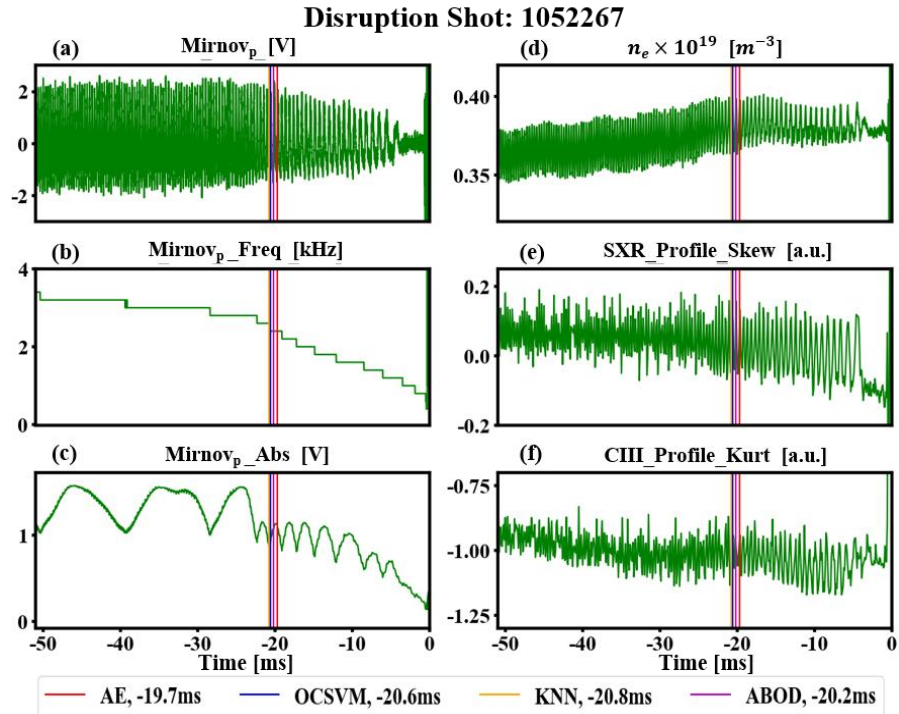

Figure 9 Precursor onset times predicted by all models on physical signals of J-TEXT, $|\Delta t|_{max} = 1.1 ms$.

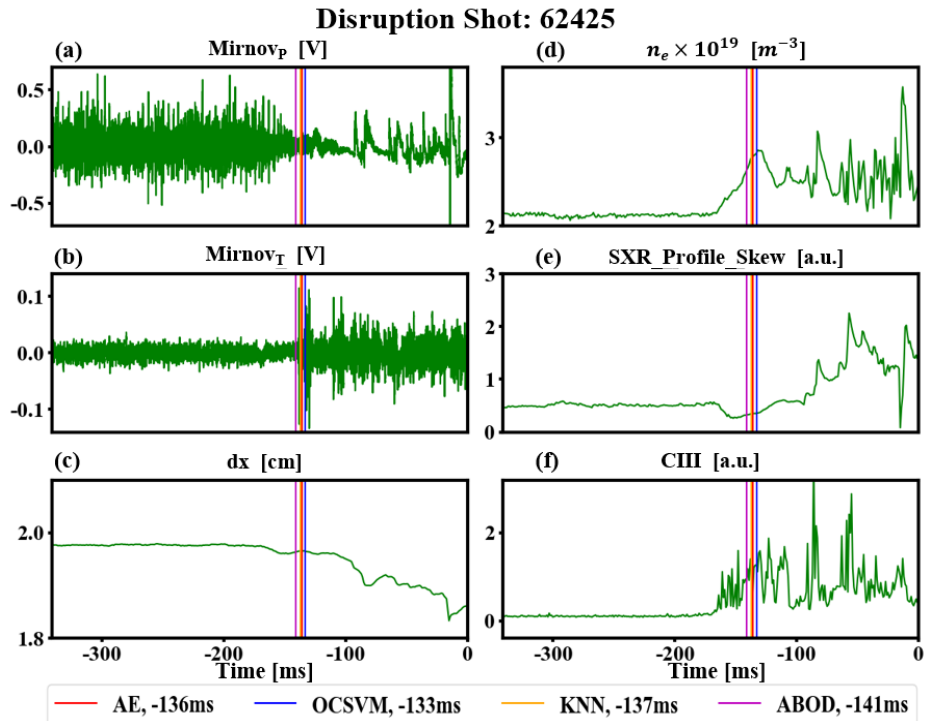

Figure 10 Precursor onset times predicted by all models on physical signals of EAST, $|\Delta t|_{max} = 8 ms$.

## 5 Precursor label Optimization

At present, the mainstream precursor labeling method for supervised learning predictor assumes that all samples in the fixed period before $t_{CQ}$ of all disruption shots in the training set are disruption precursor samples. Since the precursor periods of different disruption shots are not necessarily the same, the fixed period labeling method can result in normal samples being erroneously included in the disruption precursor sample set. This contaminated training set can have a negative impact on the performance of supervised learning predictors. In this section, the precursor onset times inferred by anomaly detection predictor are used to optimize the precursor sample labels used for training the supervised learning predictor. This labeling method is called anomaly labeling method, and the supervised learning model trained by anomaly labeling method is called the anomaly labeling model in this paper. Anomaly labeling method and fixed period labeling method are compared to explore whether labeling with anomaly detection predictor can optimize the performance of supervised learning predictor. If anomaly labeling method can optimize the performance of supervised learning model, it can prove that the disruption precursor labels obtained by anomaly detection predictor are more accurate. For this comparison anomaly labeling method is to use precursor onset time inferred by AE predictor to label positive samples. And three fixed period labeling methods are also used for comparison, which are respectively assumed that 25ms, 50ms, and 100ms before $t_{CQ}$ are the disruption precursor periods. For the same training set on J-TEXT or EAST, we assume that the longer the assumed precursor period is, the more normal samples will be mixed into the precursor sample set. So three different assumed precursor periods are selected for the fixed period labeling method to explore the relationship between model performance and contamination degree of precursor sample set. Among the three fixed period labeling methods, the 25ms labeling method has the lowest contamination degree of the precursor sample set.

Supervised learning model based on LightGBM algorithm has been established on the J-TEXT [16]. The model has very good performance and thus is used in this labeling method comparison. As LightGBM requires a balanced set of positive and negative samples, the data augmentation method is utilized to expand the positive (precursor) sample set until it reached a 1:1 ratio with the negative (normal) sample set. Then training sets obtained by different labeling methods are fed to the same supervised learning model for training. And each model is evaluated with the same test set on the same device. The minimal warning time is 10ms on J-TEXT and 20ms on EAST. The upper limit of alarm time is 200ms on J-TEXT and 300ms on EAST. The receiver operating characteristic (ROC) curves of LightGBM predictors obtained by different labeling methods on J-TEXT and EAST are shown in Figure 11, which are used to compare the performance of models. AUC is the area under curve of ROC curve. The larger the AUC, the better the overall predictor performance regarding both TPR and FPR. A perfect classifier will have a ROC AUC equal to 1. On both J-TEXT and EAST datasets, the fixed period labeling models showed a decrease in performance as the assumed precursor period increased, indicating a higher level of contamination of normal samples in the precursor sample set. Conversely, the anomaly labeling model outperformed all fixed period labeling models, demonstrating a lower degree of contamination in the precursor sample set obtained by the anomaly labeling method. These results suggest that the disruption precursor labels obtained by anomaly detection predictor are more accurate than those obtained by the fixed period labeling method.

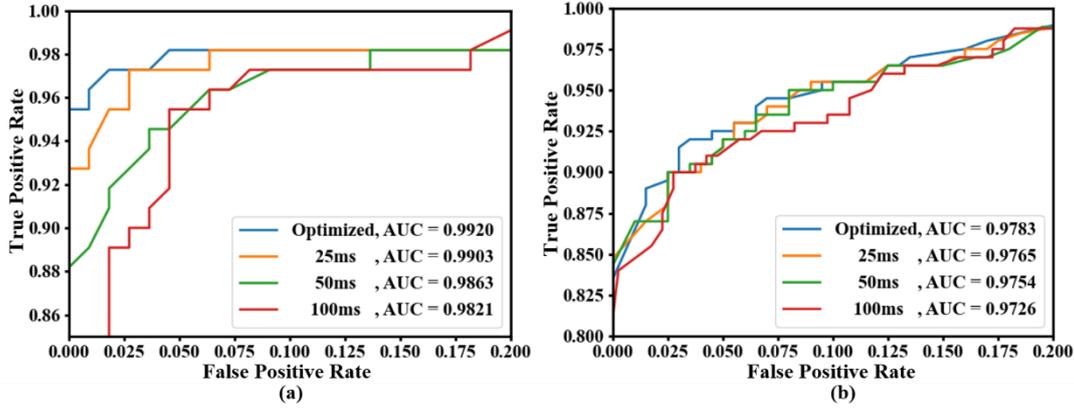

Figure 11 ROC curves of LightGBM predictors (a) on J-TEXT (b) on EAST

Furthermore, we investigated the impact of a contaminated precursor sample training set on the performance of the model. Table 8 displays the accuracy, TPR, and FPR of the predictors obtained using different labeling methods on J-TEXT. We observed that for fixed period labeling methods, an increase in the assumed precursor period results in a higher FPR for the trained model, while the TPR remains unaffected. These results indicate that if normal samples are mistakenly labeled as precursor labels in the training set, the prediction performance for non-disruptive shots will decrease, while the prediction performance for disruptive shots will be relatively unaffected. Among the models trained using different labeling methods, the anomaly labeling model exhibited the lowest FPR. This finding also validates that the contamination level of the precursor sample training set obtained using the anomaly labeling method is lower than that of fixed period labeling methods. Table 9 presents the performance of the predictor on EAST, and the results are consistent with those obtained on J-TEXT. In conclusion, the precursor onset time inferred using the anomaly detection predictor can more accurately label the precursor samples than the fixed period labeling method, thereby optimizing the performance of the supervised learning model.

Table 8 LightGBM predictor performance of different precursor labeling methods on J-TEXT

| Model | Accuracy [%] | TPR [%] | FPR [%] |
| --- | --- | --- | --- |
| Optimized | 96.82 | 97.27 | 3.64 |
| 25ms | 95.91 | 97.27 | 5.45 |
| 50ms | 94.09 | 97.27 | 9.09 |
| 100ms | 91.82 | 97.27 | 13.64 |

Table 9 LightGBM predictor performance of different precursor labeling methods on EAST

| Model | Accuracy [%] | TPR [%] | FPR [%] |
| --- | --- | --- | --- |
| Optimized | 93.75 | 92.50 | 5.00 |
| 25ms | 93.50 | 92.50 | 5.50 |
| 50ms | 93.00 | 92.50 | 6.50 |
| 100ms | 92.25 | 92.50 | 8.00 |

# 6 Summary

This study presents an investigation of the reliability of the precursor onset time predicted by the anomaly detection predictor. Four anomaly detection predictors based on Autoencoder, OCSVM, KNN, and ABOD algorithms are established. Those anomaly detection predictors based on different algorithms were tested on J-TEXT and EAST devices, and they perform well on both devices. And the inference speeds of AE, OCSVM, KNN, and ABOD predictors for a sample are compared, AE and OCSVM predictors can meet the requirements of real-time disruption prediction, while KNN and ABOD predictors cannot. Additionally, a method for inferring the disruption precursor onset time was proposed, and its reliability and accuracy were investigated. The study found that the precursor onset times are not necessarily the same for different disruption shots, and that the inferred onset times were similar for a given shot and in agreement with those recognized by human experts. These findings confirm the reliability of the anomaly detection predictors for predicting disruption precursor onset times in plasma physics experiments. Furthermore, the precursor onset times inferred by the anomaly detection predictor are utilized to optimize the precursor sample labels used for training supervised learning predictors. A comparison between the anomaly labeling method and fixed period labeling methods is made, revealing that incorrect labeling of normal samples as precursor labels leads to reduced prediction performance for non-disruption shots, while the performance for disruption shots is almost unaffected. And the performance of the anomaly labeling model is superior to that of all fixed period labeling models on both J-TEXT and EAST, indicating that the contamination degree of the precursor sample set obtained by the anomaly labeling method is the lowest. These results provide evidence that the inferred precursor onset time from the anomaly detection predictor more accurately labels precursor samples than the fixed period labeling method, leading to optimized performance of the supervised learning model. This optimization can provide higher-performance supervised learning models for future research on disruption prediction.

In the future, two studies will be conducted on J-TEXT. The first study will focus on conducting adaptive learning research on anomaly detection models, with the aim of building an adaptive model that can learn from scratch and achieve good performance on new devices using as little training data as possible. The second study will investigate how to make full use of the data from existing devices to deploy anomaly detection models on new devices with high performance, particularly in the absence of a sufficient training set at the initial stage of the new device operation. These studies will provide crucial theoretical support for future disruption prediction of the ITER reactor.

# Acknowledgments

The authors are very grateful for the help of J-TEXT team. This work is supported by the National MCF Energy R&D Program of China under Grant No.2019YFE03010004 and by the National Natural Science Foundation of China under Grant No.51821005.